# Impact of RNA Melting on Hydrating Water Structure Mapped by Femtosecond 2D-IR Spectroscopy

**Benjamin P. Fingerhut, Achintya Kundu, Jakob Schauss, and Thomas Elsaesser**
*Max-Born-Institut für Nichtlineare Optik und Kurzzeitspektroskopie, Berlin 12489, Germany.*
*Author e-mail address: fingerhut@mbi-berlin.de*

**Abstract:** We discern hydration geometries around the sugar-phosphate-backbone of an RNA double helix at the molecular level. RNA disordering upon melting is connected with a transition from ordered water structures towards local phosphate group hydration shells. © 2020 The Author(s).

## 1. Introduction

The interaction of RNA and DNA with the surrounding water environment is of fundamental relevance and has a strong impact on the macromolecular structure of the biomolecules [1-3]. Structural changes of RNA, e.g., during folding, melting and RNA translation are intimately connected to changes of hydration patterns which are not understood and barely characterized, both experimentally and theoretically. At ambient temperature and in an aqueous environment, double-stranded (ds) RNA forms an A-helix with hydrogen-bonded Watson-Crick geometries of adenine-uracil or guanine-cytosine pairs. The $(PO_2)^-$ oxygens of the sugar-phosphate-backbone are major hydration sites and point into the grooves of the RNA surface. Adjacent $(PO_2)^-$ groups can be bridged by individual water molecules and the ribose 2'-OH group can induce chain-like arrangements of hydrogen bonded water molecules in the minor groove. This water structure is expected to undergo strong changes upon melting of the A-helix, a process which we characterize for the first time at the molecular level.

## 2. Results and Discussion

The different hydration geometries and their dynamics are mapped via noninvasive vibrational probes. We demonstrate that two-dimensional infrared spectroscopy allows to follow changes of water arrangements around phosphate groups in the backbone of an oligomer dsRNA double helix dissolved in $H_2O$.

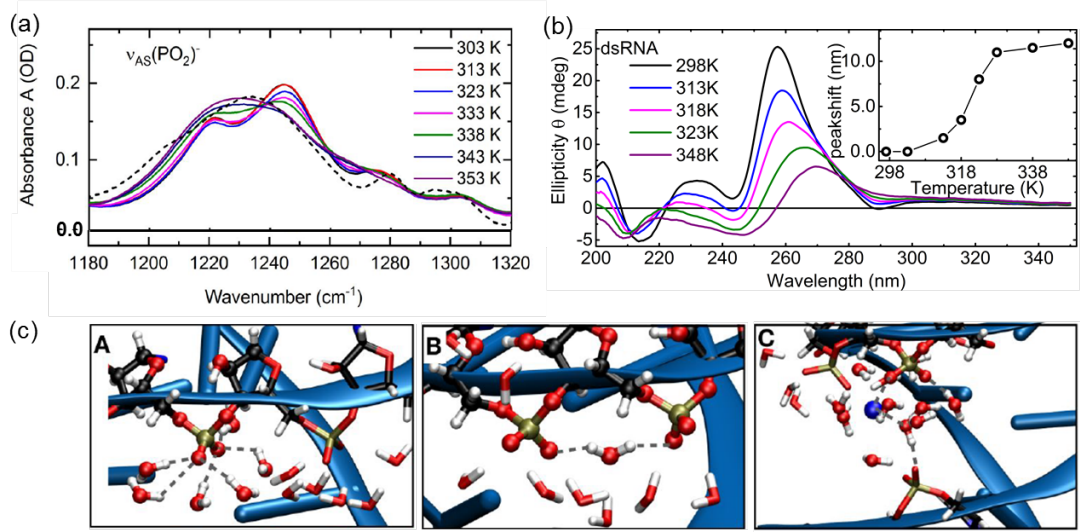

**Figure 1.** (a) Linear infrared absorption spectra of the asymmetric $(PO_2)^-$ stretching vibrations $\nu_{AS}(PO_2)^-$ of $(AU)_{23}$ RNA oligomers at different temperatures, together with the linear absorption spectrum of dsDNA at room temperature (303K, dashed black line). (b) Circular dichroism (CD) spectra of dsRNA in water at different sample temperatures $T_S$. Inset: Temperature-dependent shift $\Delta\lambda$ of the spectral maximum in CD spectra. (c) Representative snapshots of the different prototypical hydration structures around the phosphate groups of dsRNA obtained by QM/MM simulations.

Fig. 1(a) shows linear infrared absorption spectra of the asymmetric $(PO_2)^-$ stretching vibrations $\nu_{AS}(PO_2)^-$ of dsRNA oligomers (23 alternating adenine-uracil (A-U) base pairs and $K^+$ counter ions) for different sample temperatures $T_S$. The bands with maxima at 1220 and 1245 cm$^{-1}$ represent the absorption of $(PO_2)^-$ groups with different hydration environments while the weak band at 1280 cm$^{-1}$ is due to $(PO_2)^-$ - counter ion contact pairs [4]. With increasing $T_S$, the different components of $\nu_{AS}(PO_2)^-$ show substantial reshaping and the distinct absorption peaks at 1220 and 1245 cm$^{-1}$ merge into a broad line shape with a maximum around 1230 cm$^{-1}$ that closely resembles the linear infrared absorption spectra of oligomeric dsDNA (23 alternating adenine-thymine pairs, dashed line in Fig. 1a). The change of the dsRNA A-helix structure with increasing $T_S$ was followed in independent measurements of circular dichroism (Fig. 1b) and ultraviolet absorption (not shown). Upon heating, the helix undergoes a transition to a highly disordered structure with a characteristic transition temperature $T_m$=318±5 K (45±5 °C). At $T_S$=348 K, the CD spectrum of dsRNA closely resembles the CD spectrum of single-stranded RNA with identical base sequence. The absorption spectra point to the partial persistence of base stacking interactions, pointing to a highly disordered RNA structure at $T_S$>$T_m$.

Fig. 2 presents 2D-IR spectra of the RNA $\nu_{AS}(PO_2)^-$ vibration for temperatures of (a) $T_S$=303 K < $T_m$ and (b) $T_S$=348 K > $T_m$ that are compared to (c,d) 2D-IR spectra of DNA at the same sample temperatures. The absorptive 2D signal measured at a waiting time T=300 fs shows ground-state bleaching and stimulated emission due to the v=0→1 vibrational transition (yellow-red contours), and excited state contributions due to the v=1→2 absorption (blue contours). The latter are red-shifted to lower detection frequencies $\nu_3$ because of the (diagonal) anharmonicity of the vibrations (11-14 cm$^{-1}$).

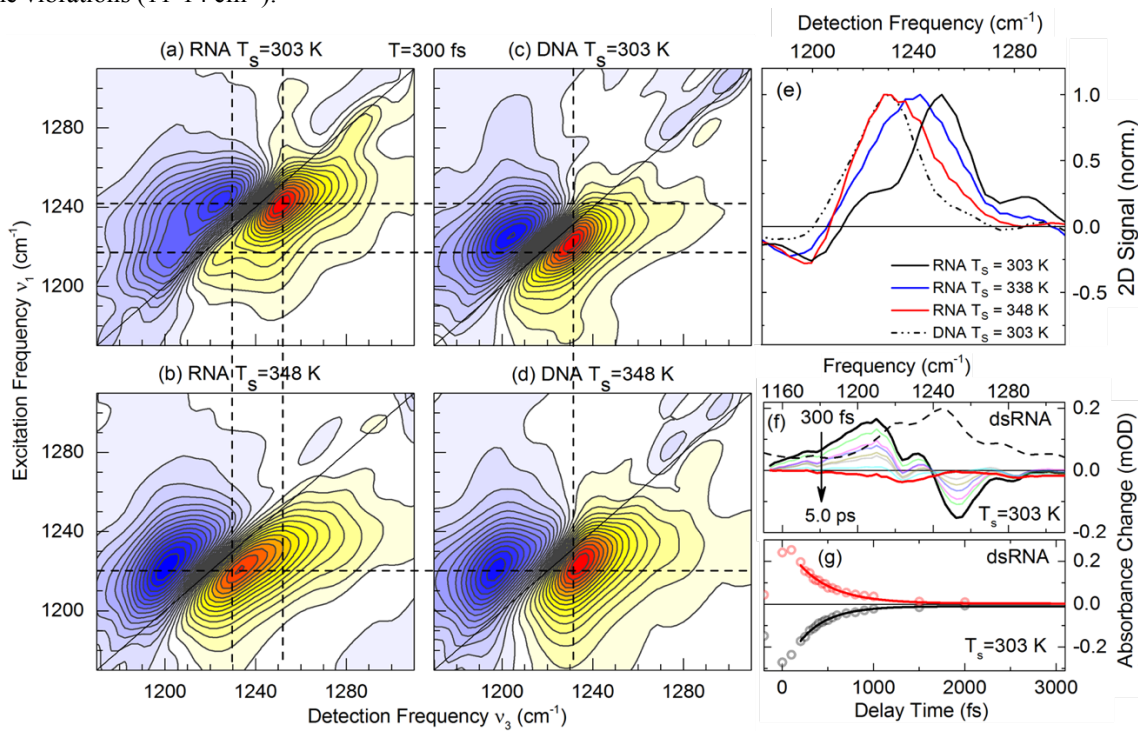

**Figure 2.** 2D-IR spectra of the asymmetric phosphate stretch vibration $\nu_{AS}(PO_2)^-$ of (a,b) dsRNA and (c,d) dsDNA recorded in water for two different sample temperatures $T_S$ = 303K and $T_S$ = 348K. The absorptive 2D spectra, i.e., the real part of the sum of the rephasing and non-rephasing photon echo signals is plotted as a function of excitation frequency $\nu_1$ and detection frequency $\nu_3$ and were measured at population time T = 300 fs. Yellow-red contours (positive 2D-IR signal) are due to bleaching and stimulated emission of the v=0→1 vibrational transition and blue contours (negative 2D-IR signal) are due to excited state absorption (v=1→2). (e) Cuts of the 2D-IR spectra along the frequency diagonal crossing the position ($\nu_1$, $\nu_3$)=(1242,1250) cm$^{-1}$, normalized 2D-IR signals are plotted as a function of detection frequency $\nu_3$. (f-h) Femtosecond pump-probe measurements of $\nu_{AS}(PO_2)^-$ of dsRNA for $T_S$ = 303K. Dashed line in panel (f): linear absorption spectrum.

At ambient temperature ($T_S$=303 K, Fig. 2a), the RNA 2D-IR spectrum displays three components at ($\nu_1$, $\nu_3$) = (1242,1250) cm$^{-1}$, ($\nu_1$, $\nu_3$) = (1220,1220) cm$^{-1}$ and ($\nu_1$, $\nu_3$) = (1283,1283) cm$^{-1}$, as clearly discerned in the diagonal cuts plotted in Fig. 2(e). The absence of cross peaks between the three contributions shows that the origin of 2D-IR signal components is from independent, uncoupled $\nu_{AS}(PO_2)^-$ vibrations. For $T_S$=348 K > $T_m$ (Fig. 2b), the main positive diagonal peak shifts to lower excitation and detection frequencies and the line shape develops into a single broad peak, inhomogeneously elongated along the diagonal $\nu_1$=$\nu_3$. The position of this peak is identical to that of the corresponding diagonal peak of DNA at $T_S=303$ K < $T_m$ (cf. dash-dotted line in Fig. 2e), with minor differences in spectral width. Notably and consistent with expectations for $(PO_2)^-/K^+$ contact ion pairs, the RNA peak at (1283,1283) cm$^{-1}$ disappears nearly completely upon heating the sample from $T_S$ = 303 to 348 K. Structural fluctuations are dominated by librational water motions occurring on a 300 fs time scale, without exchange between hydration motifs on the few-picosecond time scale of the experiments.

Population kinetics of the $\nu_{AS}(PO_2)^-$ vibration were determined in femtosecond pump-probe experiments (Fig. 2f,g). In RNA, the population lifetime has a value of 360±25 fs which is constant over the entire range of sample temperatures $T_S$ and close to the 340±20 fs lifetime of the $\nu_{AS}(PO_2)^-$ vibration of DNA [5]. The identical decay times of $\nu_{AS}(PO_2)^-$ at different sample temperatures $T_s$ result in a temperature independent lifetime broadening of the 2D lineshapes along the antidiagonals in the 2D frequency plane.

The analysis of experimental results by molecular dynamics simulations and quantum mechanical / molecular mechanical (QM/MM) calculations provides a quantitative distinction of three coexisting hydration motifs. Their relative abundance changes strongly upon RNA melting, giving rise to the changes in the 2D-IR spectra. Because of the particular sensitivity of $\nu_{AS}(PO_2)^-$ to intermolecular interactions, differences in frequency position are a direct mirror of variations in hydration structure around the phosphate groups. The component at 1220 cm$^{-1}$ (simulation: $\nu_{AS}(PO_2)^-$ = 1224.7 cm$^{-1}$, structure A in Fig. 1c) is assigned to DNA-like $(PO_2)^-$ solvation geometries with six tetrahedrally arranged water molecules in the first solvation shell around $(PO_2)^-$. For the component at 1245 cm$^{-1}$ (simulation: $\nu_{AS}(PO_2)^-$ = 1245.2 cm$^{-1}$, structure B in Fig. 1c) we find a reduced water occupancy of the first solvation shell and, in the particular realization, a water molecule bridging two adjacent phosphate groups. The solvation shells of neighboring phosphate groups are thus more ordered due to considerable overlap of hydration shells. The partial hydration by ordered water arrangements reflects the high electrostatic potential of the deep and narrow major groove and steric constraints imposed by the A-helix where the $(PO_2)^-$ oxygens point into the groove. The $\nu_{AS}(PO_2)^-$ component around 1280 cm$^{-1}$ (simulation: $\nu_{AS}(PO_2)^-$ = 1277.0 cm$^{-1}$, structure C in Fig. 1c) arises from a phosphate group in contact with a counter ion where one of the phosphate oxygen atoms replaces a water in the first solvation shell around the ion.

Overall, the dsRNA disordering upon melting induces a resorting of the relative amplitudes of the different components of $\nu_{AS}(PO_2)^-$ (Fig. 2a,b) and thus a transition from predominant ordered water structures to local DNA-like hydration shells around phosphate units. Upon heating, we further observe a substantial decrease of phosphate/K+ contact ion pairs, i.e., the reduced shielding of the repulsive Coulomb interactions between neighboring phosphate groups is connected with an increased occupancy with waters in the first hydration shell.

In conclusion, the presented results demonstrate with unprecedented detail that RNA disordering upon melting induces transitions between ordered water hydration structures and local DNA-like phosphate group hydration shells. The hydration changes reflect an increase in average numbers of $(PO_2)^-$ group hydrogen bonds that is associated with a substantial decrease of enthalpy of solvation, partially compensated by entropy changes. The resulting change of the hydration free energy during the melting process is of the same order of magnitude as free energy changes upon self-association of nucleobases [6] and thus forms a non-negligible part of the free energy balance upon RNA self-association, RNA melting and folding of complex tertiary structures.